# Truth or Square: Aspect Ratio Biases Recall of Position Encodings

Cristina R. Ceja*, Caitlyn M. McColeman*, Cindy Xiong, and Steven L. Franconeri *Member, IEEE*

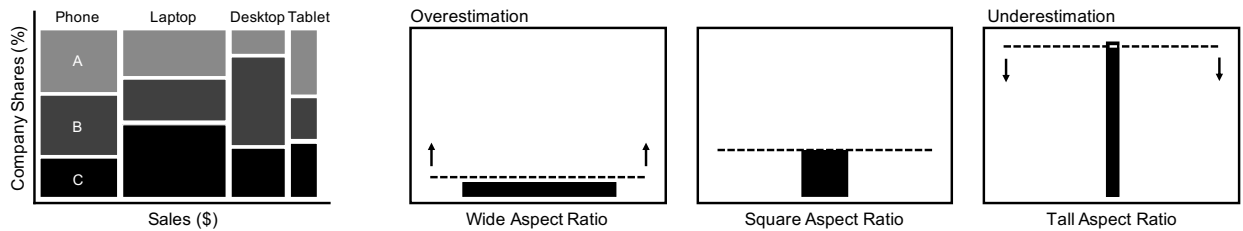

Fig. 1. The Mekko chart at left uses height to encode relative market share, similar to a bar chart. But the present results suggest that the aspect ratio of each mark may bias this height judgment. When asked to reproduce the vertical position of a single bar mark in a bar chart, bars with wide aspect ratios were overestimated, bars with tall ratios were underestimated, and bars with square ratios showed no systematic bias. This pattern of bias appeared within memory, suggesting that value comparisons that occur across time and space (e.g., bars in separate graphs) would most likely be distorted.

**Abstract**— Bar charts are among the most frequently used visualizations, in part because their position encoding leads them to convey data values precisely. Yet reproductions of single bars or groups of bars within a graph can be biased. Curiously, some previous work found that this bias resulted in an overestimation of reproduced data values, while other work found an underestimation. Across three empirical studies, we offer an explanation for these conflicting findings: this discrepancy is a consequence of the differing aspect ratios of the tested bar marks. Viewers are biased to remember a bar mark as being more similar to a prototypical square, leading to an overestimation of bars with a wide aspect ratio, and an underestimation of bars with a tall aspect ratio. Experiments 1 and 2 showed that the aspect ratio of the bar marks indeed influenced the direction of this bias. Experiment 3 confirmed that this pattern of misestimation bias was present for reproductions from memory, suggesting that this bias may arise when comparing values across sequential displays or views. We describe additional visualization designs that might be prone to this bias beyond bar charts (e.g., Mekko charts and treemaps), and speculate that other visual channels might hold similar biases toward prototypical values.

✦

## 1 INTRODUCTION

In theory, bar charts encode data values using position (defined as the vertical position of the top of a bar mark). In reality, however, an increase in the vertical position of a bar mark also results in changes to *incidental* visual properties: an increase in length, an increase in area, and a decrease in aspect ratio (width:height). Similarly, Mekko charts (see left graph in Figure 1), encode two variables separately using the width and height of a mark, but also incidentally vary in size, shape, and aspect ratio.

The present work will test whether incidental visual properties of marks, like their aspect ratio, can bias how we represent an intended encoding, using bar charts as a case study. We find that viewers recall bar positions in a biased manner, overestimating the vertical position of bar marks with wider aspect ratios, and underestimating the vertical position of bar marks with taller aspect ratios.

- *Cristina R. Ceja is with Northwestern University. E-mail: crceja@u.northwestern.edu.*
- *Caitlyn M. McColeman is with Northwestern University. E-mail: caitlyn.mccoleman@northwestern.edu.*
  *\* Both authors contributed equally to this work.*
- *Cindy Xiong is with Northwestern University and University of Massachusetts Amherst. E-mail: cxiong@u.northwestern.edu.*
- *Steven L. Franconeri is with Northwestern University. E-mail: franconeri@northwestern.edu.*

**Contributions** We document biases that arise when reproducing graphed data, using bar charts as a case study. Recent work has shown a conflicting pattern of either overestimation [36] or underestimation [23] biases for position encodings in bar charts. Yet, the causes of this bias remained unknown. In the current study, we are interested in *why* these biases occur and what underlying properties or sources might drive this bias. We seek to understand the origin of these biases, to better predict when they might arise in a variety of data visualizations, (e.g., bar charts, stacked bar charts, Mekko charts, or treemaps), that either intentionally rely on aspect ratio to encode data, or incidentally vary aspect ratio as a byproduct of their design.

## 2 RELATED WORK

### 2.1 Biases in Position Estimation

In data visualizations, quantitative values are encoded using visual encodings such as position, orientation, and saturation [4]. Some of these encoding types are less perceptually precise than others (e.g., saturation and area are encoded less precisely than position [10]), and can also be susceptible to systematic biases. For example, color hues can appear darker or lighter depending on their proximity to nearby colors [12] (see Figure 2(a)). The size of a circle will appear smaller in the context of larger circles than when surrounded by smaller circles [12] (see Figure 2(b)). One might think then that more precise forms of encoding information would be less prone to such biases. But even length, the second most precise encoding [10], can be influenced by context or orientation. Viewing a vertical line adjacent to a horizontal line of the same length makes the vertical line appear longer than the horizontal line [29] (see Figure 2(c)). Position, by some measures the most precise form of visual data encoding [10], is also not immune to biases. Two recent reports suggest that bar charts, which use position to encode data, are vulnerable to biases [23, 36]; but surprisingly, the results differ in the direction of this observed bias.

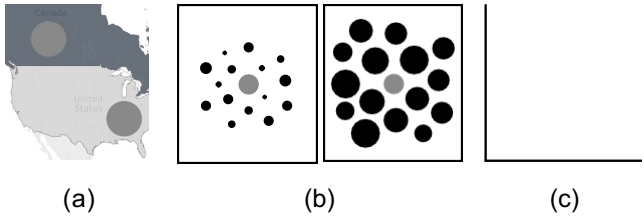

(a) (b) (c)

Fig. 2. Biases involving luminance, area, or length. (a) The gray circles are identical luminance, but the circle appears lighter when placed on a dark background *(top)* as compared to a light background *(bottom)*. (b) Both gray circles are the same size, but the gray circle in the context of the smaller, black circles *(left)* appears larger than when surrounded by larger circles *(right )*. (c) The vertical and horizontal lines are identical lengths, but the vertical line appears longer.

In a recent study by Xiong et al., viewers were asked to recreate the average position (defined as the average vertical position of the tops of the bar marks) of a group of bars with a uniform distribution (i.e., bars of the same height; see Figure 3(a)) [36]. Viewers consistently estimated the average position of the bars to be *higher* than it actually appeared. Xiong et al. claimed that this consistent overestimation indicated that our perception or memory for position encodings in bar charts is systematically biased, such that we consistently overestimate the position of bar marks. In contrast, a recent study by McColeman et al. found the reverse effect [23]. When viewers were asked to provide a position estimate for a single bar mark (see Figure 3(b)), viewers consistently estimated the vertical position of the top of the bar to be *lower* than it actually appeared.

Why did previous work on position biases find conflicting evidence? One possibility is that the tested bar marks in these studies had quite different width:height aspect ratios (see Figure 3 for an illustration), as well as different areas. In an effort to uncover why previous work found conflicting evidence for this bias [23, 36], we investigate the effect of varied aspect ratios on this vertical position estimation bias. More broadly, this work will help examine how incidental visual properties of a visualization (i.e., in the case of bar charts, aspect ratio and area) might generate biases in the recall of graphical data.

## 2.2 Impact of Aspect Ratio

Some dimensions (e.g., width, height) in visualizations are *integral*, meaning that a change in one dimension can influence how one perceives the other dimension [13]. For example, viewers are slower to verify that a stacked bar chart depicts a smaller proportion than another if the stacked bar is overall larger in size than the other bar [15].

Differences in aspect ratios can also affect performance in comparison judgments between two areas. In one study, participants were asked to compare pairs of rectangles to determine which was smaller [20]. Comparison errors were lowest when comparing pairs of rectangles with more similar aspect ratios (e.g., comparing a mark with a 3:2 ratio to one with a 2:3 ratio). In contrast, viewers made more area comparison errors when comparing pairs of rectangles with dissimilar aspect ratios (e.g., 9:1 ratio compared to a 1:9 ratio) and squares to squares (1:1 ratios). Although aspect ratio was not a direct factor in this area comparison task, it still influenced viewers' judgments of area.

The aspect ratio of graphed marks can also influence our perception of data more broadly. In line charts, distorting the aspect ratio of a slope through axis manipulation can impact the trends a viewer sees in the data and what inferences or decisions they make about it [11]. When visualizing the same dataset, extending the x-axis limits (e.g., expanding to a range of 0-100 instead of 0-50) results in a taller aspect ratio and, thus, a steeper slope in the data; shrinking the x-axis limits results in a wider aspect ratio and, thus, a more shallow slope in the data. Maintaining an aspect ratio of 1:1 (banking the line to a 45° average angle with the x-axis) seems to minimize absolute error when determining the ratio between two slopes [8, 9]. More recently, work has shown that this 45° recommendation was the result of an overly constrained data set – for more extreme slope ratio judgments, average slopes shallower than 45° result in less error for slope ratio estimations [35].

Given that aspect ratio can influence graphical perception, it is important to know when the aspect ratio of marks might bias their interpretation. Treemaps, for example, use area to encode hierarchical data, with lower levels of the hierarchy represented by smaller boxes that are contained within higher boxes which represent relatively higher levels of the hierarchy (e.g., in Figure 4, an overarching category, such as Category A, can contain multiple smaller categories, like Group A1, Group A2, etc). While area encodes these data values, the aspect ratio of each nested category mark is a byproduct of a tiling algorithm that decides where the marks should be placed in two-dimensional space, attempting to achieve an aspect ratio (width:height) closer to a square (or 1:1) ratio for each of the marks [7, 33]. However, it must trade off that preferred aspect ratio for improved spatial efficiency, resulting in marks that deviate from a square. Although the aspect ratio of each mark is incidental (that is, it is not used to directly encode data), it could potentially bias how accurately a viewer might interpret or recall the data.

Therefore, the differences in position biases found in previous work (overestimation [36] vs. underestimation [23]; see Figure 3) could have been incidentally driven by the aspect ratios of the bar marks themselves. The differing aspect ratios could have caused overestimation for reproductions of bar marks with wider aspect ratios, and underestimation for reproductions of those with tall aspect ratios. In this study, we test whether the incidental aspect ratio of the bars determines the direction of this bias.

### 2.2.1 Potential Effects of Categorical Prototypes

If aspect ratio did influence this positional bias, why did it specifically result in an overestimation for bar marks with a wider aspect ratio and underestimation for those with a taller aspect ratio? These differences in the aspect ratios of the bar marks may invite a categorical prototype effect, such that the vertical position of a bar mark is attracted toward the vertical position of a prototypical square.

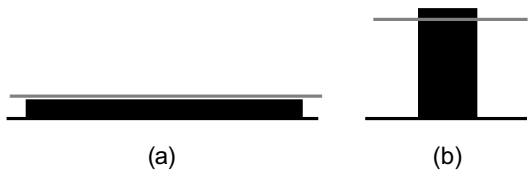

(a) (b)

Fig. 3. (a) Sample stimulus used in Xiong et al. that resulted in a systematic *overestimation* of average bar position (indicated by the *gray line*) [36]. Note that this is a group of bars, with no spacing between the individual bars. (b) Sample stimulus used in McColeman et al. which showed a systematic *underestimation* of bar position (indicated by the *gray line*) [23].

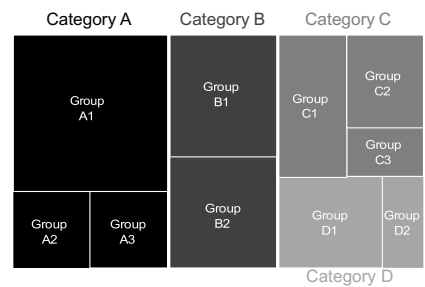

Fig. 4. A simple treemap. Each category differs in color; within each category are nested groups.

Categorical prototype effects are biases that occur when a reconstruction of a stimulus value (e.g., vertical position) must be made from an inexact representation [16, 17]. This reconstruction is influenced by both the actual stimulus value and its category prototype, resulting in a bias toward that prototype. Color, for example, has been found to be biased toward prototypical values (e.g., reporting a pinkish-red color to be more prototypically red than it actually appeared) [2,34]. In a recent study, participants were shown a color patch and asked to recreate it from memory using a color wheel after a brief delay [2]. Their responses were found to be consistent with categorical perception: color responses were attracted toward the middle of the color category and repulsed away from the category boundaries. Furthermore, when the participants were shown the same color patches but asked to respond with no time delay and with both the color patch and color wheel still present on the screen (thereby, relying less on memory), this categorical repulsion persisted.

Therefore, in Xiong et al. [36], the wide aspect ratio of that group of bars (16:1 ratio; see Figure 3(a)) might have contributed to the bias, with viewers misestimating the bars toward a more prototypical shape (i.e., a 1:1 square), resulting in overestimation. In McColeman et al. [23], however, the tall aspect ratio of a single bar mark (1:1.87 ratio; see Figure 3(b)) would have resulted in a reverse effect, with viewers underestimating the vertical position of the bar toward the vertical position of a more prototypical square.

## 3 STUDY OVERVIEW

Across three experiments, we empirically investigate the position estimation bias previously found in bar charts. Specifically, we test the role of the different incidental visual properties of these data series, and the extent to which these biases arise within memory.

Experiment 1 and 2 investigate which potential visual properties might influence the strength of this misestimation bias. Experiment 1 explores the effect of the aspect ratio of a single mark in a bar chart on its position estimation. We then untangle area from aspect ratio in Experiment 2 to investigate the extent to which area influences this position estimation bias. Finally, Experiment 3 explores memory as the potential underlying source that may contribute to this positional bias, by testing whether we *recall* position in a biased way. This design space is illustrated in Figure 5.

Our findings show that viewers are systematically biased by the aspect ratio of a bar mark when reproducing its position, such that the error and bias in position estimations were different across the tested aspect ratios. Viewers consistently *overestimated* the vertical position for marks with a *wide* aspect ratio, but consistently *underestimated* marks with a *tall* aspect ratio. Area, however, was not a contributing factor in this bias; as it was held constant across the three tested aspect ratios, area itself could not account for the observed differences in error and bias.

### 3.1 General Stimulus and Procedure

All experimental stimuli were created with MATLAB using the Psychophysics Toolbox [6, 18, 25] on an Apple Mac Mini running OS 10.10.5. The monitor was 23 inches with a 1,280 x 800 pixel resolution and a 60 Hz refresh rate. The approximate viewing distance was an average of 47 cm. All experimental materials and analyses are available at https://osf.io/nmjeq/.

Bars were generated around three set height-variant versions (low, medium, or high means) with some noise (equally sampled from +/- 0, 5, 10 pixels). The three height-variant versions introduced variability, preventing learning of the position values for any of the ratio conditions. We generated these three height-variant versions, which were each separated by 25 pixels, for each of the three ratio conditions. All ratio conditions with the medium mean contained the same mean area of 14,887px$^2$. Therefore, only trials containing bars with medium mean were analyzed, allowing us to hold area constant and isolate aspect ratio as the difference between conditions. The aspect ratios used were wide, square, or tall (see Figure 5 for examples). Bars with a wide aspect ratio had an average width:height ratio of 11.5:1 (or 414:36px), bars with a square ratio had a 1:1 ratio (or 122:122px), and bars with a tall aspect ratio had a 1:11.5 ratio (or 36:414px). All bars were 50% black (RGB: [128, 128, 128]), and presented on a 33% black (RGB: [170, 170, 170]) background. The gray shades were selected to minimize afterimages that may otherwise appear when observing high contrast images on a blank screen.

Participants first saw a stimulus display: a bar with one of three aspect ratios – wide, square, or tall – presented for 0.5 seconds near the center of the display. This was followed by a moving visual noise mask ('TV static') for 0.5 seconds to avoid potential visual afterimages that could affect participants' memory of the display. Participants then responded by dragging the top of a response probe (a bar with a wide, square, or tall aspect ratio, depending on the bar previously shown during the display) to match the vertical position of the top of the bar they had just been shown using a Mac "mighty mouse" computer mouse. The top of this response probe would randomly appear 50-80 pixels above or below the true position of the top of the bar, to avoid biases in response that may arise from consistently drawing from the top or the bottom of the stimulus. The probe would change in direct linear response to the participants' mouse movements. Response time was unlimited. After participants dragged the top of the response probe to the desired position and clicked the mouse to enter their response, another moving visual noise mask appeared, indicating the beginning of the next trial.

## 4 EXPERIMENT 1: ASPECT RATIO IMPACTS POSITION ESTIMATES

In Experiment 1, we empirically test whether bars differing in aspect ratio can influence both the accuracy and bias of position reproductions, even though aspect ratio is an incidental visual property in bar charts. Specifically, we demonstrate that viewers do not reproduce these graphed elements veridically; rather, viewers show various degrees of accuracy and bias (in the form of consistent *overestimation* for bars with wide ratios and *underestimation* for bars with tall ratios) when recalling the position of bar marks differing in aspect ratio.

### 4.1 Design and Procedure

Experiment 1 tested different aspect ratios to investigate response error and bias when redrawing these presented values. Participants performed 216 trials, during which a single bar was presented. The bar was one of a wide, square, or tall aspect ratio (see Figure 5). The aspect ratio condition (*wide, square,* or *tall*) and the height-variant versions (*high, medium,* or *low*) were all presented in random order without replacement. Each participant observed a random order of aspect ratio conditions and of height-variant versions to ensure that the variance observed in responses was not a function of order or learning effects.

Of the 216 trials, there were 72 trials each of the wide, the square, and the tall aspect ratio conditions; within each of the aspect ratio conditions, there were 24 trials each of the high, the medium, and the low height-variant versions of the display.

Twenty-five undergraduate students from Northwestern University ($M_{Age}$ = 18.72 years, $SD_{Age}$ = 1.14) participated in exchange for course credit in an introductory psychology class.

### 4.2 Results

There were nine possible conditions in each experiment, with condition as a factor defined by the aspect ratio (3 levels: wide, square, and tall) and by the height-variant versions for each aspect ratio (3 levels: low, medium, and high). Responses from only the *medium* height-variant version were analyzed, because area was held constant for these trials between the different tested aspect ratios. This decision was critical to ensure that area was not a potential driving factor of any observed biases in Experiment 1.

We prevented our formal models from overweighting extreme values within each condition by trimming the extreme 5% of values from the top and the bottom of each condition's distributions. After excluding the extreme values, there were 270 trials for each of the wide, square, and tall aspect ratios across all of the participants.

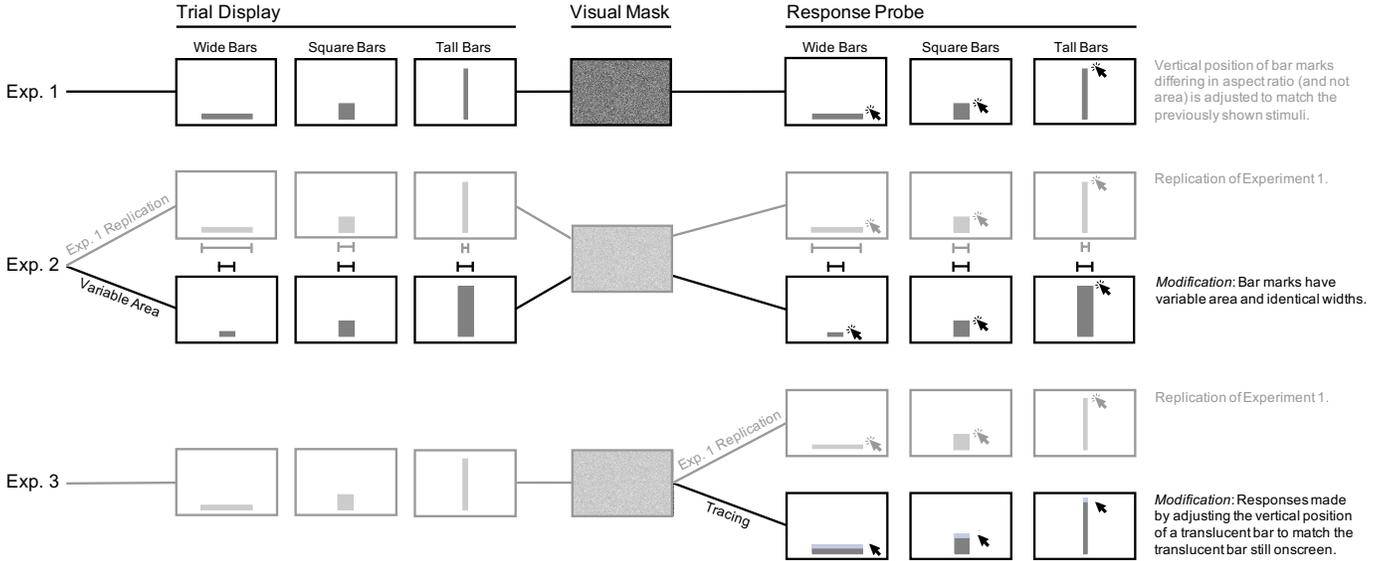

Fig. 5. Experimental procedure and design for Experiment 1, 2, and 3. All grayed boxes indicate Experiment 1 and its replication controls in Experiment 2 and 3, in which participants viewed bars of varying aspect ratios (the trial display), followed by a visual mask, and then were asked to respond by dragging the vertical position of a bar (with the same aspect ratio as the previously shown bar) to the estimated position of this previously shown bar that was no longer present on the screen; the response probe). This was then followed by another visual mask to conclude the trial (not shown here). All black boxes indicate manipulated conditions in Experiment 2 and 3, which are identical to Experiment 1 and the replication controls except for the following ways: in Experiment 2, the manipulated tracing condition differed from the replication control condition in that the widths of the bar marks were fixed across the aspect ratios, resulting in variable area between ratios. In Experiment 3, the manipulated condition involved dragging the vertical position of a translucent bar to the position of a concurrently present translucent bar (which indicated the position of the bar shown during the trial display).

Accuracy, or absolute error, was subjected to a pair of hierarchical linear mixed effects models, implemented with the LME4 package [3] in R. The first model accounted for baseline variation, with the initial bar position as a fixed effect and participant as a random effect. It was contrasted with a second model, which also included the aspect ratio of the bar marks as a fixed effect. The second model was better than the first ($\chi^2(2) = 271.6$, $p < 0.001$), reflecting a main effect of aspect ratio on absolute error. Subsequent tests on the better model did not reveal a difference between bars in the wide and square conditions ($z = 2.12$, $p = 0.086$), but did reveal a significant difference between bars in the square and tall conditions ($z = 14.46$, $p < 0.001$).

To test for differences in response bias, or signed error, we compared a base linear mixed effects model with one containing aspect ratio as a fixed effect. The model with aspect ratio as a factor was a better model overall ($\chi^2(2) = 270.03$, $p < 0.001$), indicating a main effect of aspect ratio. Subsequent pairwise tests performed on the linear mixed effects model output showed significant differences between response errors in the square and wide conditions ($z = -4.30$, $p < 0.001$), and between response errors in the tall and square conditions ($z = -12.98$, $p < 0.001$). Pairwise tests were subjected to Bonferroni correction.

To further test for response bias, we compared the *response value* with the *presented value* using paired t-tests. Significant positive t-values indicate *over*estimation; significant negative t-values indicate *under*estimation (see Table 1). We found that bars in the square condition were not significantly biased. However, bars in the wide condition were significantly overestimated, while bars in the tall condition were significantly underestimated.

Table 1. Mean differences ($\Delta_{mean}$; in pixels) between response and presented values for Experiment 1.

| Condition | $\Delta_{mean}$ | 95% Conf. | t | p |
|---|---|---|---|---|
| Wide | + 4.75 | [4.1, 5.3] | + 16.15 | <0.001 |
| Square | + 0.11 | [-0.9, 1.4] | + 0.20 | 0.842 |
| Tall | -13.86 | [-16.4, -11.3] | -10.86 | <0.001 |

### 4.3 Discussion

Experiment 1 replicates both the findings of Xiong et al., which found an overestimation bias of bar marks with a wide aspect ratio [36], and McColeman et al., which found an underestimation bias for bar marks with a tall aspect ratio [23]. While these findings initially appear contradictory, it appears that the aspect ratio of these bar marks bias position estimations in a graphical display, even though this visual property is not directly assigned to encode the data.

In addition to different trends in bias across the three aspect ratios, there also appears to be a difference in the accuracy of responses. While there was an overestimation bias for the position of bar marks with tall aspect ratios, responses in this condition were highly inaccurate and variable. In comparison, bars with square and wide aspect ratios were not nearly as prone to error as those with tall aspect ratios. Overall, this may indicate that the accuracy of responses may be influenced by the aspect ratio of the bar marks themselves, such that those with square or wide aspect ratios are more accurate than those with tall aspect ratios.

The present data also reveal that the biased representation of the three tested aspect ratios appears to be a consequence of participants misrepresenting the presented bar mark as a shape closer to a square. Viewers estimated the positions of bars with wide and tall aspect ratios in the direction of the vertical position of a more prototypical square, with the wider aspect ratios overestimated, and taller aspect ratios underestimated. Bar marks with a square aspect ratio, in comparison, were unbiased – the height and width of the bar were already equal, so a representation of a prototypical square could not further bias the perception of that mark toward the prototype.

## 5 EXPERIMENT 2: AREA DOES NOT BIAS POSITION ESTIMATES

Experiment 1 provided evidence that biases from position encodings differ based on the aspect ratio of the bar marks. Yet, aspect ratio is not the only incidental visual property typically found in bar charts. Experiment 1 held the area of each bar mark constant while varying their aspect ratios, but, in realistic displays, the area of a bar mark also increases when its vertical position increases.

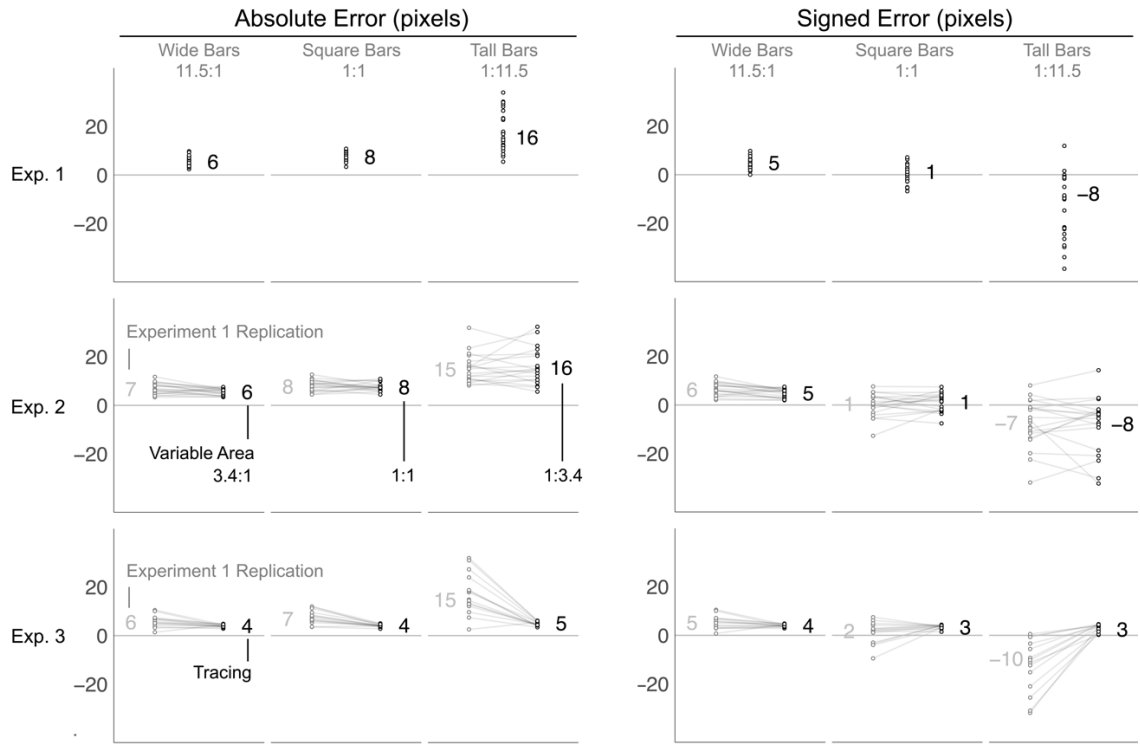

Fig. 6. Absolute error (pixels) and signed error (pixels) for Experiment 1, 2, and 3. Each dot indicates the mean error for a single participant in each corresponding experiment, and annotated numbers are the mean error in pixels for all participants. All light gray dots are for Experiment 1 or its replications in Experiments 2 and 3. All black dots are the manipulated conditions (Experiment 2: variable area condition; Experiment 3: tracing condition). The aspect ratios of the bar marks were wide (11.5:1), square (1:1), or tall (1:11.5) for all conditions but the variable area condition in Experiment 2.

Area is a salient feature for human vision. Objects are often drawn with an area proportional to their canonical or prototypical physical size [21], and irrelevant objects with greater area are salient enough to distract viewers away from smaller, target objects [26, 27]. In this experiment we investigate whether the area of the graphical bar marks biases position recall beyond the influence of aspect ratio.

Past work in perceptual psychology shows that past experience can bias estimates of area. After viewing a series of circles with varying area in a display, viewers tend to recall the size of a single circle in a way that is biased towards the average area of all of the previously seen circles [5]. In the present experiments, position estimates could be similarly compared to the average area of all bar marks across the experiment, if area were not constant between the marks. For instance, bar marks with a taller aspect ratio would have a greater area than the average area across the experiment (which would be closer to the area of the square aspect ratio bar mark). Perhaps this contrast causes viewers to somehow underestimate the height of these marks with tall aspect ratios, leading them to recall these marks as less tall a moment later (and vice-versa for lower-area wide aspect ratio marks). Bar marks with a square aspect ratio, in contrast, would have a similar area to the average area, so there would be no observed bias in position estimations.

In Experiment 2, we empirically test whether area might influence this bias in position reproductions within a bar chart. To isolate the role of area in this bias, we hold the width of the bar marks constant, resulting in different areas for the different aspect ratios, while also removing any potential effects of width on the height of the mark. As the perceived area of a rectangle can be modelled by multiplying perceived width by perceived height [1], misremembering a large width of a rectangular bar mark, for example, might inflate a viewer's vertical position estimation of the mark, resulting in overestimation. Whereas in Experiment 1 we tested three aspect ratios with area held constant, in Experiment 2 we now test three aspect ratios but with variable area (and width held constant).

### 5.1 Design and Procedure

The procedure was similar to Experiment 1, except for the addition of a *variable area* condition. The variable area condition was presented in blocks, interleaved with the *replication control* condition (a direct replication of Experiment 1). Half of the participants saw the variable area block first, while the remaining half saw the replication control block first. Counterbalancing helped to ensure that any observed differences between the variable area condition and the replication control condition were not due to learning or fatigue. In total, participants performed six blocks (3 blocks of the variable area condition, and 3 blocks of the replication control condition).

Each replication control block contained 54 trials, or 18 trials each for bars with a wide, square, and tall aspect ratio (each with 3 height-variant versions: high, medium, and low, as in Experiment 1). The variable area condition contained an identical 54 trials, with the exception that the bars now all had a width of 122 pixels (the same as the width of the square condition in all experiments). These bars, however, still contained heights matching those in the replication control blocks (wide condition: height = 36px, ratio = 3.4:1px; square condition: height = 122px, ratio: 1:1px; and tall condition: height = 414px, ratio: 1:3.4px; referred to as the *wide*, *square*, and *tall* conditions in the variable area condition, respectively). As in Experiment 1, only the *medium* height-variant versions of the wide, square, and tall conditions (in either the replication control or the variable area conditions) were included in analysis. Retaining the middle 90% of the response distribution resulted in 204 trials for analysis in each of aspect ratio conditions in both the variable area and replication control conditions.

Fifteen different undergraduate students from Northwestern University ($M_{Age}$ = 19.43, $SD_{Age}$ = 1.09) participated in exchange for course credit in an introductory psychology class.

## 5.2 Results

As in Experiment 1, participants' responses were subjected to hierarchical linear mixed effects models to test whether allowing area to vary between aspect ratios impacted estimation errors. We fit a base model with participant as a random factor and area condition (whether area was held constant, as in the replication control condition, or variable, as in the variable area condition) as a fixed factor.

We first tested for the role of area and aspect ratio on accuracy, or absolute error. The hierarchical linear model with the area condition as a factor outperformed the base model ($\chi^2(4) < 0.001$), indicating that the area condition accounted for significant variance in the model. Subsequent pairwise tests conducted on the model object did not find a significant difference (after family-wise error correction) between the wide and square conditions ($z = 2.45$, $p = 0.04$), but did find a significant difference between the square and tall conditions ($z = 4.786$, $p < 0.001$).

We used the same method to test for the influence of area and aspect ratio on bias, or signed error. An ANOVA conducted on the base model indicated that area was not a significant predictor of bias ($\chi^2 = 0.37$, $p = 0.54$). The base model was compared to a second model with aspect ratio as an additional fixed effect. The second model was a better fit to the data, indicating a main effect of aspect ratio ($\chi^2(2) = 242.95$, $p < 0.001$). Subsequent pairwise tests showed that the signed error for the tall condition was more negative than for the square condition ($z = -6.54$, $p < 0.001$). Additionally, signed error for bars in the square condition was more negative than for the wide condition ($z = -5.34$, $p < 0.001$). All pairwise tests were subjected to Bonferroni correction.

See Table 2 for a summary of the mean difference between estimated position responses and presented position values. The replication control condition followed the same pattern of results as observed in Experiment 1. Participants overestimated bars in the wide condition, responses were unbiased overall for the square condition, and participants underestimated the bars in the tall condition (see table 2). Responses, once again, displayed considerable variability (see Figure 6), especially for bars with the tall aspect ratio.

In the variable area condition, we see the same pattern with overall signed error responses. There was a slight overestimation in the wide condition ($M = 4.95$px, $SD = 4.87$), relatively little bias was found for bars in the square condition ($M = 1.29$px, $SD = 9.28$), and underestimation occurred in responses for the tall condition ($M = -7.66$px, $SD = 18.74$).

Table 2. Mean differences (in pixels) between response and presented values for Experiment 2.

| Condition | $\Delta_{mean}$ | 95% Conf. | t | p |
|---|---|---|---|---|
| Replication: Wide | + 6.44 | [5.7, 7.2] | 17.793 | <0.001 |
| Replication: Square | + 0.56 | [-0.8, 1.9] | 0.812 | 0.418 |
| Replication: Tall | - 7.20 | [-9.6, -4.8] | -5.979 | <0.001 |
| Variable Area: Wide | + 4.95 | [5, 5.6] | 14.53 | <0.001 |
| Variable Area: Square | + 1.29 | [0, 2.6] | 1.986 | 0.048 |
| Variable Area: Tall | - 7.66 | [-10.3, -5.1] | -5.842 | <0.001 |

## 5.3 Discussion

Replicating Experiment 1, Experiment 2 finds that position biases are influenced by the aspect ratio of the bar marks. This finding again provides support for previous findings of overestimation [36] and underestimation [23] for bar marks. Furthermore, we find that while aspect ratio appears to be the driving factor behind this bias, the average area of the marks does not.

Aspect ratio and area are not orthogonal dimensions in bar marks. We cannot entirely erase the effects of aspect ratio to study area in isolation, without introducing a number of extraneous variables (such as spacing and axis size), but comparing the condition where area is constant across aspect ratios (as in the replication control condition) with the condition where area varies across ratios (as in the variable area condition) does allow us to test the effects of area beyond those of aspect ratio. We found no significant change in the pattern of position bias observed when area was variable. Aspect ratio predicts bias, even when the area of the bar marks were controlled.

The accuracy (absolute error) of responses also appears to be influenced by the aspect ratio of the bar marks themselves, but not uniquely influenced by the area of the marks. Similar to Experiment 1, bar marks with square or wide aspect ratios were more accurate than those with tall aspect ratios.

Again, there also appears to be an influence of categorical prototype effects, such that the representation of these bar marks are biased by the prototype of a square. However, we find no significant difference between the bias found in the replication control and in the variable area conditions, which contained bar marks with different aspect ratios (e.g., in the replication control, bar marks in the wide condition had a ratio of 11.5:1px; in the variable area condition, these marks had a ratio of 3.4:1px). With width remaining constant across all ratios in the variable area condition, both the bars with the wide aspect ratios and tall aspect ratios were no longer as exaggerated (i.e., bars with wide aspect ratios now had an average ratio of 3.4:1px; bars with tall aspect ratios had an average ratio of 1:3.4px).

Considering that there was no significant difference in bias between the variable area and replication control conditions, these findings suggest that the direction and degree of the bias may not be influenced by the extremity of aspect ratio; otherwise we would expect more bias for the more extreme aspect ratios (i.e., for the elongated bars in the replication control condition; see Figure 5). Rather, it could be possible that there are thresholds for the aspect ratios, such that any bar mark with an aspect ratio past that threshold is biased similarly to other marks with aspect ratios beyond that threshold. Thresholds for over- and underestimation may explain why we observe the same directional trend across the discrete groups of aspect ratios (i.e., bar marks in the wide condition, regardless of their exact aspect ratios, were overestimated while those in the tall condition, again regardless of exact ratios, were underestimated). This could indicate that the reconstruction of bar marks may be more heavily biased away from the actual stimulus value by the categorical prototype, regardless of the exact aspect ratio of the mark.

## 6 EXPERIMENT 3: THE ROLE OF MEMORY IN POSITION ESTIMATES

In Experiment 3, we test how much this bias in bar marks differing in aspect ratio is driven by memory. Specifically, we test the accuracy and bias of position estimates when manipulating whether the bar stimulus is still present on the screen (when memory is no longer required to make the estimation) or when the stimulus is no longer present on the screen (requiring memory). We show that recalling the position of a bar mark from memory results in the pattern of bias found here and in previous studies of this misestimation bias [23, 36].

### 6.1 Design and Procedure

To explore the impact of memory on the observed response errors to this point, we replicated Experiment 1 (through the original *memory* condition, referred to as the *replication control* condition), and extended it to a *tracing* condition.

While the replication control condition directly mirrored that of Experiment 1, the tracing condition differed in the response phase. In this condition, participants were shown a response screen with a translucent (alpha value = 50, RGB = [128,128,128]) bar which represented the exact bar shown during the trial display. Participants were then tasked with matching the vertical position of this bar by dragging the vertical position of another equally translucent bar that was overlaid on top of the previous. The combined luminance of the two, overlaid bars during this response phase matched the luminance of the bars in Experiment 1 and 2.

This tracing response guarantees that the participants can maintain simultaneous perception of the original bar while they make their response. Having a reference probe on screen, but not centrally located with the response probe, would require eye movements or shifts of attention, which invoke memory processes. As such, in the tracing condition, both the reference probe and the response probe are centrally located to eliminate the need to transfer information from direct perception into a memory cache to provide a response.

There were six total blocks of trials, with the order of the tracing and replication control conditions counterbalanced between participants. Each block of 54 trials was broken down into 18 trials for each of the wide, the square, and the tall conditions (each with three height-variant versions: high, medium, and low). As with the previous experiments, we analyzed only the medium height-variant version for each of the wide, square, and tall conditions, and eliminated the extreme 10% of responses to avoid spurious output from our linear mixed effects models.

Twenty-one different undergraduate students from Northwestern University ($M_{Age}$ = 18.67 years, $SD_{Age}$ = 0.856) participated in exchange for course credit in an introductory psychology class.

### 6.2 Results

To test for the overall role of memory, removed from direct perception and motor error, on estimation errors, simple linear mixed effects models were constructed with participant as a random factor and with the memory/tracing conditions (whether the original stimulus was no longer present on the screen, as in the replication control condition, or remained on the screen, as in the tracing condition) as a fixed factor.

For accuracy, or absolute error, an ANOVA contrasting this base model to a model with aspect ratio added to it replicated earlier experiments: aspect ratio predicted performance ($\chi^2(4)$ = 133.97, $p <$ 0.001). Subsequent pairwise test found no significant difference between bars in the wide and square conditions ($z$ = 2.30, $p$ = 0.055), but did reveal a difference between bars in the square and tall conditions ($z$ = 5.12, $p <$ 0.001).

Testing bias, or signed error, a model with aspect ratio added as an additional fixed factor was found to be a better fit to the data: $\chi^2(4)$ = 216.5, $p <$ 0.001. The full model tested for main effects of aspect ratio ($\beta_{square}$ = -3.4, $t$ = -3.4; $\beta_{tall}$ = -14.72, $t$ = -14.8), memory/tracing ($\beta_{on}$ = -0.26, $t$ = -0.25), and interactions between aspect ratio and memory/tracing. The interactions showed relatively little difference for bars in the square condition when the original stimulus was still present on the screen versus no longer present ($\beta_{square}$ = 2.8, $t$ = 1.99), as compared to bars in the tall condition when the stimulus was still present on the screen versus no longer present ($\beta_{tall}$ = 13.27, $t$ = 9.46).

To better understand response bias, we compared the *response value* with the *presented value*, and tested for differences between these values with a set of Bonferroni-corrected t-tests (see Table 3 for complete breakdown of results).

Table 3. Mean differences (in pixels) between response and presented values for Experiment 3.

| Condition | $\Delta_{mean}$ | 95% Conf. | t | p |
|---|---|---|---|---|
| Replication: Wide | + 5.26 | [4.5, 6.0] | 14.50 | <0.001 |
| Replication: Square | + 1.86 | [0.5, 3.2] | 2.66 | 0.008 |
| Replication: Tall | - 9.53 | [-12.4 -6.6] | -6.46 | <0.001 |
| Tracing: Wide | + 4.05 | [3.8, 4.3] | 36.398 | <0.001 |
| Tracing: Square | + 3.39 | [2.9, 3.9] | 13.316 | <0.001 |
| Tracing: Tall | + 2.51 | [1.6, 3.4] | 5.459 | <0.001 |

Additionally, we tested the signed error for the tracing condition against the memory condition to see if the signed error was reliably different. The wide and square bars did not exhibit significant differences between tracing and memory conditions, $t(13)$ =< 2.00, $p >$ 0.07, but the tall bars' error was significantly lower when the bar remained on screen ($t(13)$ = 5.77, $p <$ 0.001).

### 6.3 Discussion

Experiment 3 replicated Experiment 1's findings, with position biases influenced by the aspect ratio of graphical marks. Additionally, these position biases were found to be mainly rooted in memory, as the original pattern of the biases persisted in the test of memory, but not in the test of replicating the vertical position of a bar while directly perceiving its original value.

In line with the previous experiments, we observed an overestimation of bars with wide aspect ratios, unbiased position estimation of bars with square ratios, and underestimation of bars with tall aspect ratios in the replication control condition. In the tracing condition, however, biases for all aspect ratios were instead found to be slightly *overestimated* (see Figure 6).

This small overestimation in the tracing condition may have been a consequence of motor error in an attempt to precisely trace the response probe. In Experiments 1 and 2, there was no overall bias observed when recreating the position of bars in the square condition (i.e., when the responses were made from memory); yet, only in the tracing condition of Experiment 3 did the mean signed error significantly differ from zero. We suggest that participants may have been overshooting their position responses in an effort to make sure the response probe matched the translucent bar present on the screen, not as a direct result of any perceptual biases.

This memory of the aspect ratios also seems to have been the source of the categorical prototype effects found in Experiment 1, 2, and the replication control condition in Experiment 3. The memory representation of the vertical position of the aspect ratios appears to have blended with a memory estimate of the vertical position of the categorical prototype of a square, resulting in memory biases differing in directionality based on aspect ratio.

In addition to these different trends in bias across the three aspect ratios, there once again appears to be a difference in the accuracy of responses for these aspect ratios in the replication control condition. Similar to Experiment 1 and 2, the accuracy of responses seems to be influenced by the aspect ratio of the bar marks themselves, such that bar marks with square or wide aspect ratios are more accurate than those with tall aspect ratios.

## 7 GENERAL DISCUSSION

Position is a precise way to convey data values to a viewer [10]. The present results illustrate (1) the position can be vulnerable to biases that influence how viewers recall data, and (2) the importance of not only comparing precision across encoding types, but also testing how incidental visual properties, such as aspect ratio and area, may be influential byproducts of an explicit encoding like position.

To align the variance observed between the current conditions and these earlier studies, we compare our observed bias against a scaled version of the error reported in [10] by dividing the signed error by the height of the bar, such that the error is scaled to the bar height. The resulting proportion corrected bias (*signederror/barheight*) can be compared against the transformed value from a replication of the visual variable precision ranking experiment [10, 14]). The difference in precision between position encodings and length encodings was approximately 4% in the earlier work. Our conditions yield a similar scale of error: the signed error as a proportion of bar height differs by 2% between bars with a square aspect ratio and those with a tall ratio. This suggests that the bias effect found here has around half the impact of the choice of encoding dimension.

## 8 LIMITATIONS, FUTURE DIRECTIONS, AND POTENTIAL GUIDELINES

**Aspect Ratios Bias on a Continuum, or a Spectrum?** The present study was motivated to resolve seemingly conflicting findings on vertical position estimates in bar charts from previous studies [23, 36], which explored *discrete* aspect ratios (wide and tall, respectively). Similarly, our investigation of discrete aspect ratios found that the position of bar marks with wide aspect ratios tends to be overestimated while the position of those with tall ratios is underestimated. However, it is still unclear whether this position estimation bias is linearly proportional to aspect ratio, or if it changes qualitatively across the different possible aspect ratios: this bias may follow a discrete pattern separated by a threshold, such that once the aspect ratio of a bar mark passes a certain threshold, position estimation is biased by a similar amount to any other aspect ratios past that threshold. Future work could expand the range of aspect ratios examined to determine whether this bias is truly discrete, or is either a simple continuous (linear) function or a complex continuous function (e.g. exhibiting categorical biases) [23].

**Aspect Ratio in Chart Design** While the present study focused on exploring the effect of aspect ratio on vertical position estimation with *individual bars*, future research could investigate how *relative* aspect ratios could influence vertical position recall in *a set of bars*. Past research has demonstrated that adjacent shapes can distort the representation of a target shape [32], suggesting that if a viewer is unable to fully isolate a single mark, they may be susceptible to integrating the values of surrounding, irrelevant data points into their representation of that mark. This may result in the overall average width and height of all of the marks' various aspect ratios impacting the representation of the single, chosen mark.

**Other Data Encodings** Recall of position appears to be influenced by categorical prototype effects, such that the position of a mark is misestimated towards a position that would render a more categorical shape, such as a square. But the vertical position of bar marks may not be the only visual encoding that is susceptible to categorical prototype effects. Category boundaries between hues, for example, can bias how one perceives or remembers colors toward the center of color categories [2,34]. Therefore, data values encoded with rainbow colormaps might be biased toward a prototypical color category [31], as they use hue encodings which can be subject to prototype effects. Similarly, a size legend illustrating examples of small and large values, could create categorical prototypes, biasing viewers to see medium-large dots as larger, or small-medium dots as smaller.

**Differentiating Memory and Perception** In Experiments 1 and 2 of the present study, participants responded from memory during the position estimation task, as the bar mark shown during the trial was no longer present on the screen. Yet, even though the response must rely heavily on memory to recall position, the observed response also incorporates both perception (through *viewing* the shown bar) and motor response (through *replicating* the shown bar with the mouse). Any observed error could be attributable to any of those three components (memory, perception, or motor error).

Based on the present study, we are unable to make conclusive claims about how strongly perception contributes to this bias, beyond the influence of memory. The output of the reproduction task we used in the study can speak only to the fidelity of the reproduction between the encoding and the execution; it cannot speak to any perceptual error that may warp the encoding (input) or the representation because the perceptual error would be shared by both the input and the output.

The El Greco fallacy illustrates this inability to test perception in reproduction-like tasks [30]. The artist El Greco was thought to have astigmatism because he often painted elongated figures. Yet, astigmatism could not have caused the elongation of the art because, if he had a pervasive vision problem, then his canvas should have also appeared elongated to him. Because the same (mis)perception should have applied to both the figure and the surface it was reproduced on, his painting style should have self-corrected, such that the figure would be drawn to normal proportions. The same fallacy prevents the current reproduction method for position estimates from uncovering potential 'online' biases in perception, because those same biases should also be present in the reproductions.

Isolating the stage of this position bias would help mitigate its effects, as each source could involve different design prescriptions. If this effect is truly limited to memory, then visualization researchers would need to leverage techniques (e.g., captions or focusing techniques [19]) to reinforce take-away messages or precise data value reading to prevent biases in position recall. In contrast, if this effect might actually stem from perception, then a different set of design prescriptions (e.g., unifying chart aspect ratios in dashboards and small multiples [28]) would need to be made to mitigate biased perception.

**Reporting Methods** In this current work, we investigated the visual representation of position using a visual reporting method (i.e., visually recreating the vertical position of the bar mark). Yet, visual images have been shown to be potentially dual-coded, such that these images are both visually and verbally encoded [24]. For example, when we view a data visualization, we not only have a visual representation, but also a verbal representation, which may convey semantic information about it (e.g., there was an increasing trend, the average position was high, etc.). Previous work has also shown that verbal labels for an image can influence its memory representation, with this representation being pulled towards a more prototypical representation of the image [22]. Since we only tested visual representations with a visual reporting method in the present work, it is unknown whether the verbal representations of a visual encoding within a visualization are also biased. Therefore, future work should explore whether this bias also appears for other ways of reporting data values, such as verbal reports (e.g., estimate the y-value of the shown bar by reporting a number from 1-100) or a two-alternative forced choice task (e.g., which of two charts has a mark with a higher data value?).

## 9 CONCLUSION

The present clarifies previous conflicting findings of an overestimation [36] and underestimation [23] bias for position in bar charts, by showing that these biases are driven by an incidental visual property of the visualization: the aspect ratio of the bar mark. Although both area and aspect ratio are byproducts of the design of a bar chart, only different aspect ratios, not different areas, elicit varying degrees of accuracy and bias in position estimates of bars. The findings from this study indicate that our representation of the position of graphical marks can be biased by incidental visual properties, as we find that recall of position encodings are influenced by the aspect ratio of the bar present in the graph.

## 10 ACKNOWLEDGMENTS

This work was supported in part by the National Science Foundation Graduate Research Fellowship under grant No. DGE-1842165, and in part by grant IIS-1901485 from the National Science Foundation.

The authors wish to thank Gabriel Belkind, Maksim Giljen, Victoria Kam, Jun Hwa Lee, and Chase Stokes for assistance in data collection, and Evan Anderson for helpful comments.

## REFERENCES

[1] D. Algom, Y. Wolf, and B. Bergman. Integration of stimulus dimensions in perception and memory: Composition rules and psychophysical relations. *Journal of Experimental Psychology: General*, 114(4):451, 1985.

[2] G.-Y. Bae, M. Olkkonen, S. R. Allred, and J. I. Flombaum. Why some colors appear more memorable than others: A model combining categories and particulars in color working memory. *Journal of Experimental Psychology: General*, 144(4):744–763, 2015.

[3] D. Bates, M. Mächler, B. Bolker, and S. Walker. Fitting Linear Mixed-Effects Models Using lme4. *Journal of Statistical Software*, 67(1):1–48, 2015.

[4] J. Bertin, W. J. Berg, and H. Wainer. *Semiology of Graphics: Diagrams, Networks, Maps*, volume 1. University of Wisconsin Press: Madison, 1983.

[5] T. F. Brady and G. A. Alvarez. Hierarchical encoding in visual working memory: Ensemble statistics bias memory for individual items. *Psychological Science*, 22(3):384–392, 2011.

[6] D. H. Brainard and S. Vision. The psychophysics toolbox. *Spatial Vision*, 10:433–436, 1997.

[7] M. Bruls, K. Huizing, and J. J. Van Wijk. Squarified treemaps. In *Data Visualization 2000*, pages 33–42. Springer, 2000.

[8] W. S. Cleveland. *Visualizing Data*. Hobart Press, 1993.

[9] W. S. Cleveland, M. E. McGill, and R. McGill. The shape parameter of a two-variable graph. *Journal of the American Statistical Association*, 83(402):289–300, 1988.

[10] W. S. Cleveland and R. McGill. Graphical Perception: Theory, Experimentation, and Application to the Development of Graphical Methods. *Journal of the American Statistical Association*, 79(387):531–554, 1984.

[11] M. Correll, E. Bertini, and S. Franconeri. Truncating the y-axis: Threat or menace? In *Proceedings of the 2020 CHI Conference on Human Factors in Computing Systems*, pages 1–12, 2020.
[12] M. Fineman. *The nature of visual illusion*. Courier Corporation, 2012.
[13] W. R. Garner. *The processing of information and structure*. Psychology Press, 2014.
[14] J. Heer and M. Agrawala. Multi-scale banking to 45 degrees. *IEEE Transactions on Visualization and Computer Graphics*, 12(5):701–708, 2006.
[15] J. Hollands and I. Spence. Integral and separable dimensions in graph reading. In *Proceedings of the Human Factors and Ergonomics Society Annual Meeting*, volume 41, pages 1352–1356. SAGE Publications Sage CA: Los Angeles, CA, 1997.
[16] J. Huttenlocher, L. V. Hedges, and S. Duncan. Categories and particulars: Prototype effects in estimating spatial location. *Psychological Review*, 98(3):352, 1991.
[17] J. Huttenlocher, L. V. Hedges, and J. L. Vevea. Why do categories affect stimulus judgment? *Journal of Experimental Psychology: General*, 129(2):220, 2000.
[18] M. Kleiner, D. Brainard, D. Pelli, A. Ingling, R. Murray, C. Broussard, et al. What's new in Psychtoolbox-3. *Perception*, 36(14):1, 2007.
[19] C. N. Knaflic. *Storytelling with data: A data visualization guide for business professionals*. John Wiley & Sons, 2015.
[20] N. Kong, J. Heer, and M. Agrawala. Perceptual guidelines for creating rectangular treemaps. *IEEE Transactions on Visualization and Computer Graphics*, 16(6):990–998, 2010.
[21] T. Konkle and A. Oliva. Canonical visual size for real-world objects. *Journal of Experimental Psychology: Human Perception and Performance*, 37(1):23, 2011.
[22] G. Lupyan. From Chair to "Chair": A Representational Shift Account of Object Labeling Effects on Memory. *Journal of Experimental Psychology: General*, 137(2):348, 2008.
[23] C. McColeman, M. Feng, L. Harrison, and S. Franconeri. No mark is an island: Precision and category repulsion biases in data reproductions. *IEEE Transactions on Visualization and Computer Graphics*, In Print.
[24] A. Paivio. *Imagery and Verbal Processes*. Psychology Press, 2013.
[25] D. G. Pelli and S. Vision. The VideoToolbox software for visual psychophysics: Transforming numbers into movies. *Spatial Vision*, 10:437–442, 1997.
[26] M. J. Proulx. Size matters: large objects capture attention in visual search. *PloS One*, 5(12), 2010.
[27] M. J. Proulx and H. E. Egeth. Biased competition and visual search: the role of luminance and size contrast. *Psychological Research*, 72(1):106–113, 2008.
[28] Z. Qu and J. Hullman. Keeping multiple views consistent: Constraints, validations, and exceptions in visualization authoring. *IEEE Transactions on Visualization and Computer Graphics*, 24(1):468–477, 2017.
[29] J. O. Robinson. *The Psychology of Visual Illusion*. Courier Corp, 1998.
[30] I. Rock. *The nature of perceptual adaptation*. Basic Books, 1966.
[31] B. E. Rogowitz and L. A. Treinish. Data visualization: the end of the rainbow. *IEEE Spectrum*, 35(12):52–59, 1998.
[32] K. B. Schloss, F. C. Fortenbaugh, and S. E. Palmer. The configural shape illusion. *Journal of Vision*, 14(8):23–23, 2014.
[33] B. Shneiderman and M. Wattenberg. Ordered treemap layouts. In *IEEE Symposium on Information Visualization, 2001. INFOVIS 2001.*, pages 73–78. IEEE, 2001.
[34] D. A. Szafir. The good, the bad, and the biased: Five ways visualizations can mislead (and how to fix them). *Interactions*, 25(4):26–33, 2018.
[35] J. Talbot, J. Gerth, and P. Hanrahan. An empirical model of slope ratio comparisons. *IEEE Transactions on Visualization and Computer Graphics*, 18(12):2613–2620, 2012.
[36] C. Xiong, C. R. Ceja, C. J. Ludwig, and S. Franconeri. Biased Average Position Estimates in Line and Bar Graphs: Underestimation, Overestimation, and Perceptual Pull. *IEEE Transactions on Visualization and Computer Graphics*, 26(1):301–310, 2019.